\documentclass[12pt]{article}
\usepackage{graphicx} 
\usepackage[toc,title,page]{appendix}
\usepackage{cite} 
\usepackage{amsfonts}
\usepackage{amsmath}
\usepackage{amssymb}
\usepackage{physics}
\usepackage[dvipsnames]{xcolor}
\usepackage{feynmp-auto}
\usepackage{epsfig}
\usepackage{slashed}
\allowdisplaybreaks
\usepackage{comment}
\usepackage{authblk}
\usepackage{blindtext}
\usepackage{hyperref}
\usepackage{cleveref}
\usepackage{setspace}
\usepackage{tikz}
\usepackage{lipsum}
\usepackage{ragged2e}
\usepackage{xcolor}
\usepackage{soul}
\usepackage[c]{esvect}
\usepackage{harpoon}
\usepackage{subcaption}
\usetikzlibrary{angles,quotes}
\usepackage[english]{babel}
\usepackage[autostyle]{csquotes}
\usepackage{nicematrix}
\usepackage{blkarray}

\newcommand{\bc}{\begin{center}}
\newcommand{\ec}{\end{center}}

\renewcommand{\bf}[1]{\mathbf{#1}}

\newcommand{\ua}{\uparrow}
\newcommand{\da}{\downarrow}
\newcommand{\ra}{\rangle}
\newcommand{\la}{\langle}
\newcommand{\nn}{\nonumber}
\newcommand{\mt}{\mathcal{T}}

\title{ Regularized Entanglement Entropy of Electron-Positron Scattering with a Witness Photon }
\author{Shanmuka Shivashankara$^{1,2}$\footnote{sshivash@brown.edu}, Grace Gogliettino$^2$\footnote{ggogliettino@colgate.edu} }
\affil{\itshape\small $^1$ Brown University, \itshape\small Providence, RI\ 02912  USA
}\affil{\itshape\small $^2$ Colgate University, \itshape\small Hamilton, NY\ 13346  USA
}
\date{
\small }

\numberwithin{equation}{section}

\begin{document}

\maketitle
\begin{abstract}
Regularized quantum information metrics are calculated for the scattering process $e^-e^+ \rightarrow \gamma,Z\rightarrow \mu^-\mu^+$ that has a witness photon entangled with the initial electron-positron state.  Unitarity implies the correct regularization of divergences that appear in both the final density matrix and von Neumann entanglement entropies.  The entropies are found to quantify uncertainty or randomness.  The variation of information, entanglement entropy, and correlation between the muon's and witness photon's helicities are found to convey equivalent information.  The magnitude of the muon's expected helicity rises (falls) as the helicity entropy falls (rises).  Area, or the scattering cross section, is a source of entropy for the muon's helicity entropy and momentum entropy.  The muon's differential angular entropy distribution is similar to the differential angular cross section distribution, capturing the forward-backward asymmetry at high center of mass energies.
\end{abstract}


\section{Introduction}

Recently, unitarity is upheld in \cite{shiva1}-\cite{shiva2} while calculating quantum information science (QIS) metrics such as entropy for scattering and decay processes at tree level.  They find starkly different results than published works \cite{Seki}-\cite{Araujo2}.  These latter works dropped unitarity by ignoring the forward scattering amplitude when considering scalar and electromagnetic interactions.  Unitarity affects the QIS calculations in a variety of ways.  By keeping unitarity,  both the normalization constant and the total von Neumann entanglement entropy remain unchanged after a decay or scattering process.  Not keeping unitarity implies incorrect physical consequences.  e.g., works \cite{Araujo1,Araujo2} investigated the QIS metrics for a witness particle while ignoring unitarity.  A witness or spectator particle is entangled with scattering particles but does not participate in the interaction.  They found that the entanglement entropy of the witness is altered by the interaction.  However, \cite{shiva1} provides a proof that unitarity requires the witness particle's reduced density matrix and entanglement entropy remain unchanged after the interaction.

Another property of unitarity is the appropriate regularization of final density matrices.  After tracing over initial particles while keeping unitarity, the final density matrix will consist of a direct sum of two terms.  The first term is the probability for no interaction to occur.  The second term pertains to the interaction having occurred with degrees of freedom such as spin and momenta specifying the final states.  By keeping unitarity when calculating the final density matrix of a polarized muon decay process, the authors in \cite{shiva2} find the probability for the decay not to occur is $1-\Gamma T$ where $\Gamma$ is the total decay width of the parent particle and $T=2\pi \delta (0)$ is the unregularized time.  Setting this latter probability to zero gives both the conditional density matrix for the decay to occur and the regularization $T=1/\Gamma$.  Hence, the regularized $T$ is the lifetime of the parent particle.  Without unitarity, $\Gamma T$ would not appear in the final density matrix \cite{Seki}-\cite{Araujo2}.  Confirmation of this regularization is had by calculating both the correct expected neutrino helicity and trace of a density matrix, which should be $-1$ and $1$, respectively.  Whereas \cite{shiva2} uses the $S$-matrix to calculate the entropy for a decay process, the Wigner-Weisskopf method yields a similar result with a much longer and difficult calculation \cite{lello}.  With the above regularization, both finite density matrices and a finite mutual information amongst the degrees of freedom can be calculated.  Awareness of calculating finite density matrices and QIS metrics in field-theoretic interactions is lacking in the literature.  A review of QIS for particle physicists can be found in \cite{Lykken}.

Herein, an electron-positron scattering process, $e^-e^+\rightarrow \gamma Z  \rightarrow \mu^-\mu^+ $, is considered with a witness photon.  The witness photon is entangled with the electron-positron pair but does not partake in the interaction.  This process merits study because similar processes have been studied without imposing both unitarity and regularization.  In section \ref{fdensity}, unlike recent works \cite{Seki}-\cite{Araujo2}, unitarity is maintained up to tree level while calculating the final density matrix.  Also, unlike these prior works, a $Z$ boson propagator is included.  Setting to zero the probability for the scattering not to occur will give the appropriate regularization.  This regularization equals the total $accessible\ area$ for scattering.  A regularization ansatz for the final density matrix of a general decay or scattering process is provided in equation (\ref{genreg}).  Due to regularization, the QIS calculations herein are $finite$ unlike recent works in the literature. 

In section \ref{QIS}, finite reduced density matrices and finite von Neumann entropies for the muon's momentum and helicity and witness photon's helicity are derived.  A variety of QIS metrics are plotted versus the center of mass energy.  The entanglement entropy, variation of information, and correlation of helicities between the muon and witness photon convey similar information. Entropy rises (falls) when the magnitude of the correlation falls (rises).  The graphs of the muon's expected helicity and helicity-entropy show dramatic behavior when the $Z$-boson is on-shell.  As the magnitude of the expected helicity falls (rises), the helicity entropy rises (falls).  Area, momentum, and quantum mechanics are found to be the three sources of momentum entropy, which is the information-theoretic Sackur-Tetrode equation.  Akin to the muon's differential angular cross section distribution, the differential angular $entropy$ distribution is symmetric at low energies and shows a forward-backward scattering asymmetry at higher energies due to the $Z$-boson.  The latter entropy distribution is inverted with respect to the cross section distribution.  These results suggest that the von Neumann entanglement entropy quantifies randomness.

\section{Density matrix regularization for an electron-positron scattering with a witness photon}\label{fdensity}

\begin{figure}[ht]
\bc
\includegraphics[width=.6\textwidth]{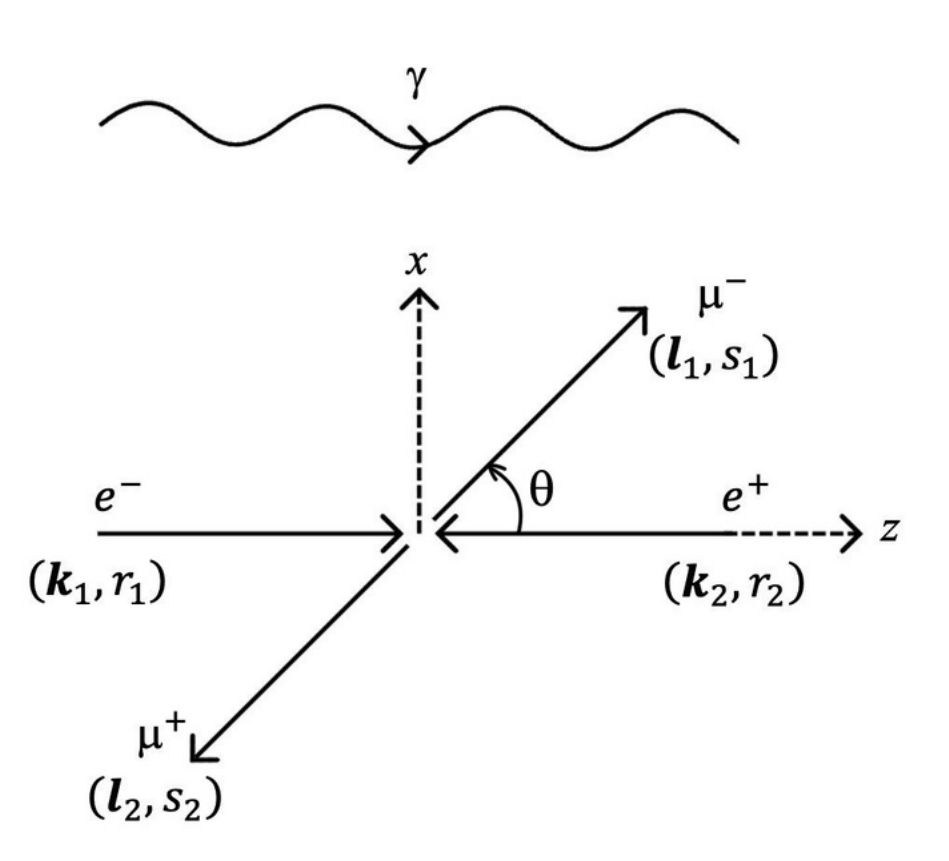}
\caption{ The scattering process $e^- e^+ \rightarrow \mu^- \mu^+$ occurs in the presence of a witness photon, $\gamma$, that is entangled with the electron-positron pair. The momentum-helicity pairs for the electron and positron are  $(\vb*k_i, r_i), i=1,2$.  The momentum-helicity pairs for the muon and antimuon are  $(\vb*l_i, s_i), i=1,2$.}
 \label{scattering}
\ec
\end{figure}

Consider an electron-positron scattering process with a witness photon (see Figure \ref{scattering}).  The photon is a $witness$, i.e. entangled with the initial electron-positron state, but not a participant in the electron-positron scattering. The objective is to calculate the regularized final density matrix for the witness photon ($\gamma$) and final state muon-antimuon pair ($\mu^-,\mu^+$).  With that matrix, a variety of QIS metrics are calculated in section \ref{QIS}.

The final state will have fermion-antifermioin pairs but not include $ZH$, $\gamma \gamma$, Bremsstrahlung, etc.  Due to tracing out particles and the optical theorem, the inclusion of these missing final states does not alter the reduced density matrix of $\mu^-,\ \gamma$.  Therefore, apart from the witness photon, the final state has a Fock space of two particles since the electron and positron either do not interact or scatter into a final pair of particles.  A free 2-particle Hamiltonian has a basis that spans the final state, implying a direct sum of final Hilbert spaces $\mathcal{H}_{e^-} \otimes \mathcal{H}_{e^+} \oplus\sum\limits_x\mathcal{H}_x \otimes \mathcal{H}_{\bar x}$.   A two-particle state is written as
$|\vb*l_1,s_1;\vb*l_2,s_2\rangle = |\vb*l_1,s_1\rangle\otimes |\vb*l_2,s_2\rangle,
$ where $(\vb*l_i,s_i), i=1,2$ are the momentum-helicity pairs for the two particles. The inner product of a state is
$\langle \vb*{p},r|\vb*{q},s\rangle$=
$2E_{\vb*{p}} (2\pi)^3\delta^{(3)}(\vb*p-\vb*q)\delta^{r,s}.$   

Assume the initial (pure) state is a superposition of the electron, positron, and a witness photon. 
\begin{align}\label{initial}
|i\rangle = \cos\alpha|RL ; \da \rangle + \sin \alpha\ e^{i\beta}|LR ; \ua \rangle
\end{align}
e.g., $|RL;\da \rangle \equiv |RL\rangle_{e^-e^+}\otimes|\da \rangle_\gamma$ represents a right-handed electron, left-handed positron, and a spin down or left-handed witness photon.  Leptons are assumed massless.  Each possible initial state has a zero net angular momentum.  The momenta are suppressed.  $\alpha$ and $e^{i\beta}$ are the entanglement parameter and relative phase, respectively.  The initial density matrix is 
$\rho^i=|i\rangle\langle i|.$  Assume the initial state is normalized or Tr$\rho^i = 1$.
$\gamma$'s reduced density matrix of its polarizations is \begin{align}\label{redg}\rho^i_\gamma= \cos^{2}\alpha|\da \rangle\langle \da|+\sin^{2}\alpha| \ua \rangle\langle \ua|
\end{align}
and is unchanged by the interaction because of unitarity \cite{shiva1}.  After the electron-positron annihilation, the final state is $|f\rangle=S|i\rangle=(1+i\mt)|i\ra$, where $\mt$ and $S$ are the transition operator and unitary $S$-matrix, respectively.  

Inserting the identity operator $1 =  \sum_{x} I_{x \bar x}$ for possible final 2-particle states ($x,\bar x$), the final state has the form
\begin{align*}
|f\rangle =|i\rangle + (\sum_{x} I_{x \bar x})\ i\mt |i\rangle.
\end{align*}
($x, \bar x$) above pertains to a lepton-antilepton pair or a quark-antiquark pair. 

The operator $I_{x\bar x}$ above has the form $\prod\limits_{i=1,2} \ Q_{\vb*l_i,s_i} |\vb*l_{i}s_{i}\rangle\langle \vb*l_{i}s_{i}| \equiv \\\prod\limits_{i=1,2}\ \sum\limits_{s_i}\int \dfrac{d^{3}\vb*l_{i}}{2E_{\bf{\vb*l}_i} (2\pi)^3}|\vb*l_{i}s_{i}\rangle\langle \vb*l_{i}s_{i}|$.  After scattering,
the final density matrix is $\rho^{f}=|f\ra\la f|$ or
\begin{align}\label{rf}
\rho^f=&\ |i\rangle\langle i|+ \sum_{x} I_{x \bar x}\ (i \mt)|i\rangle\langle i|+\text{h.c.}\nn \\
&+ \sum_{x} I_{x \bar x}\ (i\mt)|i\rangle\langle i|(-i\mt^\dagger)\ \sum_{x} I_{x \bar x}.
\end{align}
The last term in the above equation gives the possible final states along with coherence terms between different particles.  The first three terms pertain to no interaction having occurred.

Tracing over all three initial particles and using the optical theorem \cite{peskin}, the first three terms in equation \ref{rf} give the probability that no scattering  occurs between the initial electron and positron, i.e.  
\begin{align*}
\la i | i \ra + \la i|i\mt|i\ra + \la i|(-i\mt^\dagger)|i\ra &= 1 - \frac{\cos^2\alpha \sum_f \sigma_{RL \rightarrow f}+\sin^2 \alpha \sum_f \sigma_{LR \rightarrow f}}{V/(2T)}.
\end{align*}
In the references \cite{Seki}-\cite{Araujo2}, the above forward scattering amplitudes are missing.  In the denominator above, $V = (2\pi)^3 \delta^3(0)$ and  $T = 2\pi \delta(0)$ are the unregularized volume and time, respectively.  In an arbitrary reference frame,  $V/(2T)$ above becomes $V/(\upsilon_{12}T)$, where $\vb*\upsilon_{12}$ is the relative velocity of initial particles. In the center of mass frame, the relative velocity is two times the velocity of light since the electron-positron pair is massless.  $V/(\upsilon_{12}T)$ represents the accessible area for the interaction. The appropriate regularization for the above divergence, $V/(2T)$, is had by setting the above probability for $no\ scattering$ to zero or
\begin{align}\label{reg}
V/(2T) &\equiv \cos^2\alpha \sum_f \sigma_{RL \rightarrow f}+\sin^2 \alpha \sum_f \sigma_{LR \rightarrow f}=\sigma_T.
\end{align}
 This gives the conditional density matrix for an interaction to occur.  Equation (\ref{reg}) states that the total cross-section ($\sigma_{T}$) times the luminosity ($\upsilon_{12}/V$) is the scattering rate (1/T) (cf. in \S 116 in \cite{landau}).  Confirmation of this regularization is had by calculating the trace of the final density matrix, expected helicity, and expected momentum of $\mu^-$ at the second to last paragraph of section \ref{QIS}. 

A recipe for regularizing the final density matrix for any interaction can be written succinctly.  Assume the initial density matrix was not normalized.  Then, the Lorentz invariant regularization under the $S$-matrix formalism is the ansatz  
\begin{align}\label{genreg}
\dfrac{\la i|i \mt| i\ra + \la i|(i \mt)^\dagger| i\ra}{\la i|i\ra} = -1,
\end{align}
where 
$\la i|i\ra = \prod\limits_{j=1}^n (2E_jV)$.

$n$ above is the number of initial particles.  $E_j$ and $V$ are the initial energy of the $j$th particle and unregularized volume, respectively.  State $|i\ra$ represents the initial state of particles.  For an initial state given by equation (\ref{initial}) where $n=3$, equation (\ref{genreg}) implies equation (\ref{reg}) as the correct regularization.  If $n=1$, the interaction is a decay with the regularization $T=2\pi \delta(0) = \dfrac{1}{\Gamma_{Total}}$, where $\Gamma_{Total}$ is the total decay width of the initial particle \cite{shiva2}.  If $n=2$, there might exist a scattering process between two particles or a decay process with a witness.  For $n=4$, pairs of particles partake in scattering with possible initial entanglement.      
 
 If the initial state in equation (\ref{initial}) does not have a witness photon, equation (\ref{reg}) acquires an interference between the two possible initial states.  The interference would increase the total scattering cross-section for an electromagnetic interaction.  Therefore, the presence of a witness photon erases the interference.    

\begin{figure}[htp]
\subfloat[]{
\includegraphics[clip,width=.85\columnwidth]{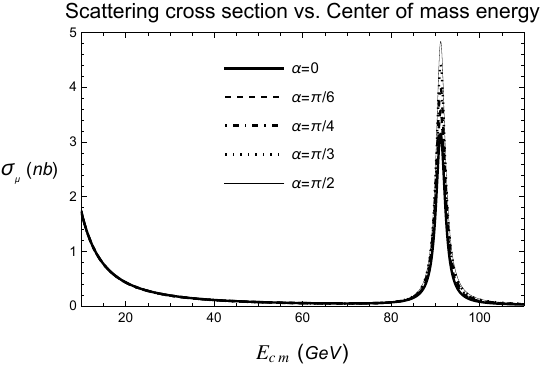}%
}

\subfloat[]{
\includegraphics[clip,width=.85\columnwidth]{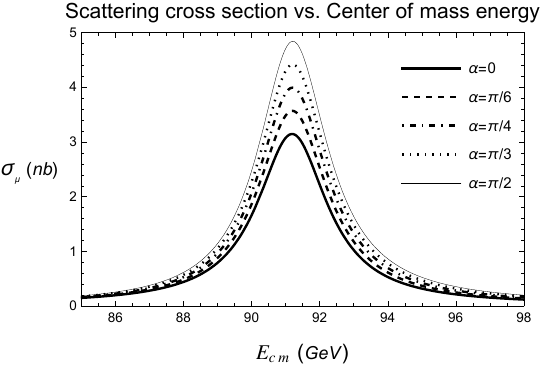}%
}
\caption{Cross section for the process $e^-e^+\rightarrow \gamma, Z \rightarrow \mu^-\mu^+$ in nanobarns ($nb$).  (a) At low energy, the curves are nearly identical for the different entanglement parameters, $\alpha$.  (b) Near the $Z$-boson mass, the curves separate.}
\label{sigma}
\end{figure}

The tree level electron-positron scattering process is mediated by a Higgs boson, a photon, and $Z$ boson since the final states are lepton-antilepton pairs or quark-antiquark pairs.  The Higgs channel is small and ignored assuming the center of mass ($c.m.$) energy is not near the Higgs mass.  Suppose the scattering was only electromagnetic.  Owing to electromagnetic interactions being parity invariant, equation (\ref{reg}) becomes $V/(2T) = \sum_f\sigma_{RL \rightarrow f}= \sum_f\sigma_{LR \rightarrow f}$ and the entanglement parameter $\alpha$ would be lost in the regularization.  Figure \ref{sigma} shows that the scattering cross section for the process $e^-e^+\rightarrow\mu^-\mu^+$ is pronounced at the $Z$-boson mass, where  its cross section is 
\begin{align}\label{simgamu}
\sigma_\mu \equiv \cos^2\alpha \ \sigma_{RL \rightarrow \mu^-\mu^+} + \sin^2 \alpha \ \sigma_{LR \rightarrow \mu^-\mu^+}.    
\end{align}
As $\alpha$ approaches $\pi/2$ in Figure \ref{sigma}(b), the cross section peaks since the $Z$-boson channel weighs the left-handed coupling more so than the right-handed coupling.

\section{Reduced density matrices and QIS metrics for the muon and witness photon}\label{QIS}

Tracing equation (\ref{rf}) over all particles except the witness photon's polarizations and $\mu^-\mu^+$ provides the normalized reduced density matrix.
\begin{align}\label{rhomumug}
\rho^{f}_{\mu^-\mu^+\gamma} &= \cos^2\alpha| \da\rangle\langle \da|\ 
\bf{A} + \sin^2\alpha| \ua \rangle\langle \ua |\ 
\bf{B} + \cos\alpha\sin\alpha\ e^{-i\beta}|\da \rangle\langle \ua |\ \bf{C}\nn \\
&\quad + \text{h.c.},
\end{align}
where
\begin{align*}
&\bf{A} = 1 - \dfrac{1}{V/(2T)}\  \sigma_{RL \rightarrow \mu^- \mu^+} + \frac{1}{(2EV)^2}\ I_{\mu^-\mu^+}(i\mt)|RL \rangle\langle RL|(-i\mt^\dagger)I_{\mu^-\mu^+},\\
&\bf{B} = 1 - \dfrac{1}{V/(2T)}\  \sigma_{LR \rightarrow \mu^- \mu^+}  + \frac{1}{(2EV)^2}\ I_{\mu^-\mu^+}(i\mt)|LR \rangle\langle LR|(-i\mt^\dagger)I_{\mu^-\mu^+},\\
&\bf{C} = -\frac{1}{V/(2T)}\ \frac{1}{2(2E)^2}\ \Big (\prod_{i=1,2} Q_{\vb*l_is_i}\Big) \mathcal{M}^{\vb*l_1,s_1;\ \vb*l_2,s_2}_{\vb*k_1,R;\ \vb*k_2,L }\Big(\mathcal{M}^{\vb*l_1 s_1;\ \vb*l_2 s_2}_{\vb*k_1,L;\ \vb*k_2,R}\Big)^\dagger * \\
&\quad\quad (2\pi)^4\delta^{4}(\sum l_i-\sum k_i) +\frac{1}{(2EV)^2}\ I_{\mu^-\mu^+}(i\mt)|RL\rangle\langle LR |(-i\mt^\dagger)I_{\mu^-\mu^+}.
\end{align*}
$\mathcal{M}^{\vb*l_1,s_1;\ \vb*l_2,s_2}_{\vb*k_1,r_1;\ \vb*k_2,r_2 }$ above is the Feynman scattering amplitude given by 
\begin{align*}
i\mathcal{M}^{\vb*l_1,s_1;\ \vb*l_2,s_2}_{\vb*k_1,r_1;\ \vb*k_2,r_2 }=(2\pi)^4\delta^{(4)}(\sum_i k_i - \sum_i l_i)\ \la \vb*l_1,s_1;\ \vb*l_2,s_2|i\mathcal{T}| \vb*k_1,r_1;\ \vb*k_2,r_2\ra.   
\end{align*}

The second term in both $\bf{A}$ and $\bf{B}$ derive from the optical theorem as does the first term in $\bf{C}$.  Notice that tracing over $\mu^-$ and $\mu^+$ will make $\bf{C}$ exactly zero.  $\rho^f_{\mu^-\mu^+\gamma}$ is a block diagonal matrix.  One matrix is 2 by 2 and pertains to the witness photon's polarizations.  The second matrix is infinite dimensional with degrees of freedom in momenta and polarizations of $\mu^-\mu^+$ and also polarizations of the witness photon $\gamma$.  The Feynman scattering amplitudes and relevant numerical constants are given in appendix \ref{feynamp-appendix}. 

Letting $V/(2T)\rightarrow \sigma_T$ (see equation (\ref{reg})), tracing over $\mu^+$ and the muon's momentum ($\vb*l_1$) gives the normalized two-particle reduced density matrix as a direct sum.  

\begin{align}\label{rhomug}
&
\begin{blockarray}{ccc}
 & |\da \ra &  |\ua \ra   \\
\begin{block}{c(cc)}
  \la \da| & \cos^2\alpha\ (1 - \dfrac{\sigma_{RL \rightarrow \mu^- \mu^+}}{\sigma_T}) & -\dfrac{ \sin2\alpha\  e^{-i\beta}}{2\sigma_T}\ \sigma_{RLLR\rightarrow \mu^- \mu^+}  \\
  \la \ua| & -\dfrac{ \sin2\alpha\  e^{i\beta}}{2\sigma_T}\ \sigma_{LRRL\rightarrow \mu^- \mu^+} & \sin^2\alpha\ (1 - \dfrac{\sigma_{LR \rightarrow \mu^- \mu^+}}{\sigma_T})\nn  \\
\end{block}
\end{blockarray}\\
 & \quad \quad \oplus\ \dfrac{\sigma_\mu}{\sigma_T} \rho^f_{\mu^-\gamma}
\end{align}
$\sigma_{LRRL\rightarrow \mu^-\mu^+}$ is the $mixed$ scattering cross section wherein the two Feynman amplitudes are opposites in handedness as pertains to the electron and positron. 

Equation (\ref{rhomug}) is a direct sum of matrices.  The first matrix refers to no $\mu^-\mu^+$ final state.  Its trace is the probability for no $\mu^-\mu^+$ final state, $1-\frac{\sigma_\mu}{\sigma_T}$.  The second matrix or second term in the direct sum is the probability ($\frac{\sigma_\mu}{\sigma_T}$) for a $\mu^-\mu^+$ final state times the density matrix $\rho^f_{\mu^-\gamma}$ assuming $\mu^-\mu^+$ is the final state.  $\rho^f_{\mu^-\gamma}$ has a trace of one and is given by
\begin{align}\label{rhoh}
\rho^f_{\mu^-\gamma}  =& \dfrac{1}{\sigma_\mu} \sum_{s_1t_1}\int\frac{d^3\vb*l_1}{(2\pi)^32E_{\vb*l_1}} \Big(\cos^2\alpha\ A^{t_1s_1}_{\vb*l_1RLRL}\ |\da \rangle\langle \da |\nn+ \sin^2\alpha\   A^{t_1s_1}_{\vb*l_1LRLR}\ |\ua \rangle\langle \ua |\nn \\
&+ \frac{\sin2\alpha\  e^{-i\beta}}{2}\ A^{t_1s_1}_{\vb*l_1 RLLR}\ |\da \rangle\langle \ua | + h.c.\Big)\ | t_1 \rangle \langle s_1 |,
\end{align}
where 
\begin{align}\label{Al1}
A^{t_1s_1}_{\vb*l_1 abcd}=&\frac{1}{2(2E)^2}\sum\limits_{s_2}\int\frac{d^3\vb*l_2}{(2\pi)^32E_{\vb*l_2}}\mathcal{M}^{\vb*l_1,t_1;\vb*l_2,s_2}_{\vb*k_1,a;\vb*k_2,b}\Big(\mathcal{M}^{\vb*l_1 s_1;\vb*l_2 s_2}_{\vb*k_1,c;\vb*k_2,d}\Big)^\dagger *\nn\\
&(2\pi)^4\delta^{(4)}(\sum l_i - \sum k_i)
\end{align}
with $a$ and $c$ ($b$ and $d$) being the handedness $R,L$ of the $e^-$ ($e^+$).\\

$\rho^f_{\mu^-\gamma}$ above has information about $\mu^-$'s polarization ($t_1,s_1$) along with the witness photon's polarization.  It provides information about correlation between the photon and $\mu^-$.  Tracing over $\mu^-$ gives the same $initial$ reduced density matrix for the witness photon's polarizations (see equation (\ref{redg})).  In other words, unitarity does not allow the scattering to affect the entropy of the witness photon \cite{shiva1}.  If unitarity was not kept, both the change in the witness' reduced density matrix and change in entanglement entropy would not be zero (cf. \cite{Araujo1, Araujo2}).  Appendix \ref{rhomug-appendix} gives the $4x4$ matrix $\rho^f_{\mu^-\gamma}$.  

\begin{figure}[htp]
\subfloat[]{
\includegraphics[clip,width=\columnwidth]{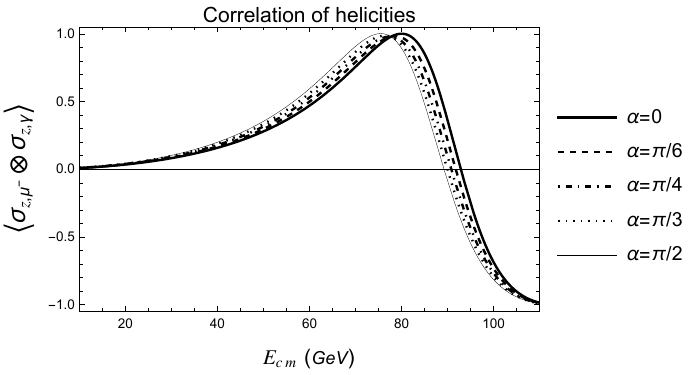}%
}

\subfloat[]{
\includegraphics[clip,
width=
\columnwidth]{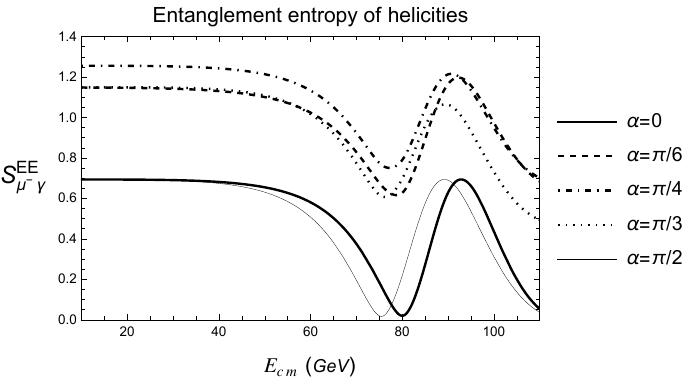}%
}
\caption{(a) and (b) compare the correlation and entanglement entropy of helicities between the muon ($\mu^-$) and the witness photon ($\gamma$) for different entanglement parameters ($\alpha$).  Correlation is plotted in units of $\hbar^2/2$.  The magnitude of the correlation is inversely related to the entropy.}
\label{helicities}
\end{figure}
 
With $\rho^f_{\mu^-\gamma}$, the correlation and entropy of helicities can be plotted with respect to the $c.m.$ energy.   The correlation of helicities is given by $\la \sigma_{z,\mu^-}\otimes \sigma_{z,\gamma}\ra = tr(\sigma_{z,\mu^-}\otimes \sigma_{z,\gamma}\ \rho^f_{\mu^-\gamma})$, where $\sigma_{z,\mu^-}= \sigma_{z,\gamma}$ is the diagonal Pauli spin matrix.  The entanglement entropy is given by $S^{EE}_{\mu^-\gamma}=-tr(D^f_{\mu^-\gamma} \log D^f_{\mu^-\gamma})$, where $D^f_{\mu^-\gamma}$ is the diagonal matrix of $\rho^f_{\mu^-\gamma}$. Figure \ref{helicities} compares the correlation and entanglement entropy of helicities between the muon and witness photon for different entanglement parameters, $\alpha$. The peaks of the correlation of helicities correspond to entanglement entropy minima.  Zero correlation corresponds to entropy maxima.  As the magnitude of the correlation rises, the entropy falls.  Other correlations ($\la \sigma_{x,\mu^-} \otimes \sigma_{z,\gamma} \ra = \la  \sigma_{y,\mu^-} \otimes \sigma_{z,\gamma} \ra$) are zero. $\sigma_{x,\mu^-}$ and $\sigma_{y,\mu^-}$ are the two off-diagonal Pauli matrices 

Tracing equation (\ref{rhomumug}) over $\mu^+$ and the polarizations of the muon and witness photon gives the following continuous reduced density matrix of the muon's momentum. 
\begin{align*}
& 1-\dfrac{\sigma_\mu}{\sigma_T}\ \oplus\ \dfrac{\sigma_\mu}{\sigma_T} \rho^f_{\vb*l_1}\nn
\end{align*}
$\rho^f_{\vb*l_1}$ in the last term above is the density matrix of the muon's momentum assuming $\mu^-\mu^+$ is a final state. It is given by
\begin{align}\label{rhomom}
\rho^f_{\vb*l_1} = \Big(\dfrac{V}{(2\pi\hbar)^3}\int d^3\vb*l_1\Big)\ \sum_{s_1}\dfrac{f(\vb*l_1)_{s_1s_1}}{V} \ \dfrac{|\vb*l_1\rangle\langle\vb*l_1|}{2E_{\vb*l_1}V}, 
\end{align}
where
\begin{align*}
f(\vb*l_1)_{t_1s_1} = \dfrac{(\hbar c)^5}{c}\dfrac{1}{\sigma_\mu\ 2E_{\vb*l_1}}\ \ (\cos^2\alpha\ A^{t_1s_1}_{\vb*l_1RLRL}+ \sin^2\alpha\   A^{t_1s_1}_{\vb*l_1LRLR}). 
\end{align*}
In equation (\ref{rhomom}) above, the continuous sum over states $(\dfrac{V}{(2\pi\hbar)^3}\int d^3\vb*l_1)$ is the analogue of a discrete sum ($\sum\limits_n$) over states.  The dimensionless volume ratio $\sum\limits_{s_1}\dfrac{f(\vb*l_1)_{s_1 s_1}}{V}$ occupies the normalized matrix position specified by $\dfrac{|\vb*l_1\rangle\langle\vb*l_1|}{2E_{\vb*l_1}V}$.  $\hbar$ and $c$ are inserted for numerical calculations.

The von Neumann entanglement entropy of the muon's momentum given a $\mu^-\mu^+$ final state is 
\begin{align}\label{sl1}
S^{EE}_{\vb*l_1} =& -tr(\rho^f_{\vb*l_1}\log\rho^f_{\vb*l_1})\nn\\
=& -\dfrac{V}{(2\pi\hbar)^3}\int d^3\vb*l_1\  \Big(\sum\limits_{s_1}\dfrac{f(\vb*l_1)_{s_1s_1}}{V}\Big) \log (\sum\limits_{s_1}\dfrac{f(\vb*l_1)_{s_1s_1}}{V}) \nn\\
=& -4\log \hbar + 2\log \sigma_\mu + \int d\cos\theta\ \dfrac{1}{\sigma_\mu}\dfrac{d\sigma_\mu}{d\cos\theta} \log (\dfrac{1}{\mathcal{M(\cos\theta)}}),
\end{align}
where $\mathcal{M(\cos\theta)} \equiv \frac{1}{4} \sum \limits_{s_1s_2}\dfrac{\cos^2\alpha |\mathcal{M}^{s_1 s_2}_{RL}|^2 +  \sin^2\alpha |\mathcal{M}^{s_1 s_2}_{LR}|^2}{s^2/c^4}$ and $s$ is the electron-positron $c.m.$ energy squared.   In the above $information-theoretic$ Sackur-Tetrode equation, there are three sources of entropy or uncertainty: quantum mechanics ($\hbar$), area ($\sigma_\mu$), and momentum($\mathcal{M(\cos\theta)}^{-1/4}$).  This is similar to a recent work on a polarized muon decay, wherein the daughter electron's momentum entropy has the following form (see equation (3.4) in reference \cite{shiva2}).  
{\begin{align}\label{st2}
S^{EE}_e = -2\log \hbar + \log \frac{V}{2cT} + \int d\Omega \int dE\ \dfrac{1}{\Gamma}\dfrac{d^2\Gamma}{d\Omega dE} \log (\dfrac{4 E}{f(E,\theta)c^2})
\end{align}
$T$ and $\Gamma$ above are the muon's lifetime and decay width, respectively.  $\theta$ is the electron's scattering angle with respect to the muon's spin. $E$ is the electron's energy.  Unlike equation (\ref{sl1}), equation (\ref{st2}) has an unregularized volume $V$.

\begin{figure}[htp]
\subfloat[]{
\includegraphics[clip,width=\columnwidth]{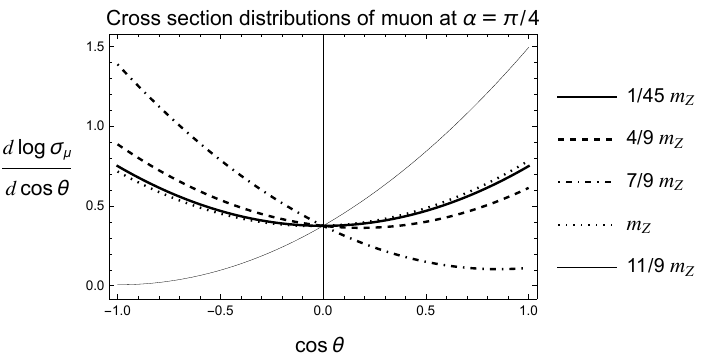}%
}

\subfloat[]{
\includegraphics[clip,width=
\columnwidth]{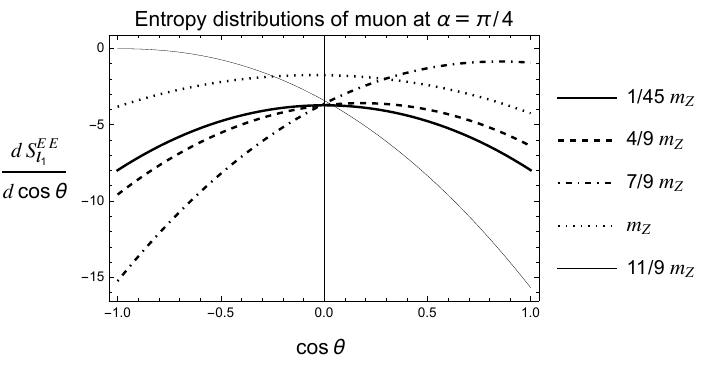}%
}
\caption{The differential cross section and entropy angular distributions of the muon are compared for different $c.m.$ energies at an entanglement of $\alpha=\pi/4$.  $\theta$ is the muon's scattering angle with respect to the incident electron.  Both distributions are peaked in absolute terms at back scattering (forward scattering), for high $c.m.$ energies but below (above) the $Z$-boson mass, $m_Z$.  At low $c.m.$ energy, both distributions are symmetric.}
\label{figangular}
\end{figure}

The (weighted) angular cross section distribution $\frac{d \log \sigma_\mu}{d \cos \theta} = \frac{1}{\sigma_\mu}\frac{d\sigma_\mu}{d\cos\theta}$ and the angular entropy distribution  $\frac{d S^{EE}_{\vb*l_1}}{d \cos \theta}$ (see equation (\ref{sl1})) are plotted in Figure \ref{figangular}.  Both distributions are symmetric in the muon's scattering angle at low $c.m.$ energies,  e.g. about $2\ GeV$ or $\frac{1}{45}m_Z$, where $m_Z$ is the $Z$-boson's mass.  At higher energies, the forward-backward scattering asymmetry is conveyed by both distributions.  As the $c.m.$ energy crosses $m_Z$, the distributions go from being peaked at backward scattering to being peaked at forward scattering in absolute terms.      

Tracing equation (\ref{rhomug}) over the witness photon's polarizations gives the following reduced density matrix of the muon's polarization.
\begin{align*}
1-\dfrac{\sigma_\mu}{\sigma_T}\ \oplus\ \dfrac{\sigma_\mu}{\sigma_T}\rho^f_\lambda\nn
\end{align*}
$\rho^f_\lambda$ above is the density matrix of the muon's helicity ($\lambda$) assuming $\mu^-\mu^+$ is the final state. It is given by
\begin{align}\label{rhogam}
\rho^f_\lambda=\sum\limits_{s_1t_1}\Big(\dfrac{V}{(2\pi\hbar)^3}\int d^3\vb*l_1\ \dfrac{f(\vb*l_1)_{t_1s_1}}{V} \Big) \ |t_1\rangle\langle s_1|. 
\end{align}
If the leptons have mass, the above density matrix will have coherence terms.  Therefore, when calculating the von Neumann entanglement entropy ($S^{EE}_{\lambda}$) of the muon's polarization, the matrix $\rho^f_\lambda$ would be diagonalized first.  This very diagonalization makes irrelevant the sign ambiguity of the Feynman amplitudes or coherence terms. For the massless case,
\begin{align}\label{slambda}
S^{EE}_{\lambda}=&-tr(\rho^f_\lambda \log \rho^f_\lambda)\nn \\
=&-\sum_{s_1}\Big(\dfrac{V}{(2\pi\hbar)^3}\int d^3\vb*l_1 \ \dfrac{f(\vb*l_1)_{s_1s_1}}{V}\Big ) \log\Big(\dfrac{V}{(2\pi\hbar)^3}\int d^3\vb*l_1 \ \dfrac{f(\vb*l_1)_{s_1s_1}}{V}\Big ),\nn \\
&
\end{align}
where $f(\vb*l_1)_{s_1s_1}$ is given below equation (\ref{rhomom}).  This article assumes massless leptons.  Area, or the scattering cross section ($\sigma_\mu$), is a source of uncertainty or entropy in the helicity entropy, $S^{EE}_\lambda$.

\begin{figure}[htp]
\subfloat[]{
\includegraphics[clip,width=\columnwidth]{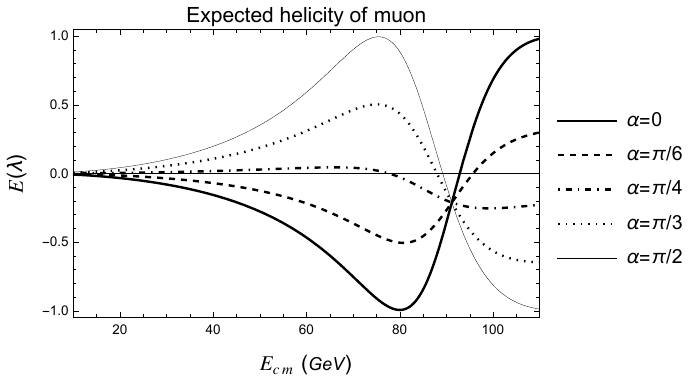}%
}

\subfloat[]{
\includegraphics[clip,width=\columnwidth]{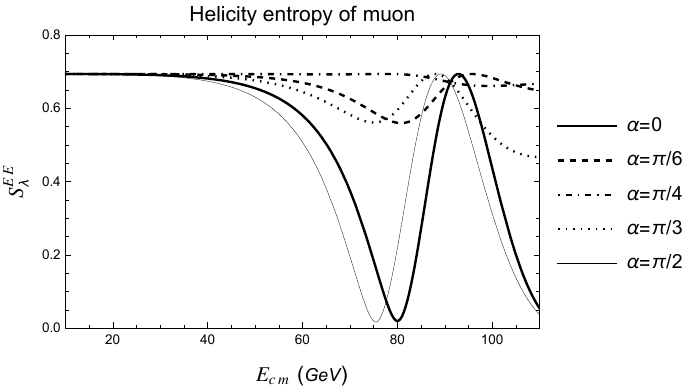}%
}
\caption{(a) The expected helicity, $E(\lambda)$, of the muon is plotted in units of $\hbar/2$ versus the center of mass energy, $E_{cm}$.  $\alpha$ is the entanglement parameter.  $E(\lambda)$ is $-.21$ when the $Z$-boson is on-shell. (b) The associated von Neumann entropy of the muon's helicity is $S^{EE}_\lambda=.67$ when the $Z$-boson is on-shell.}
\label{fighelicity}
\end{figure}

\begin{figure}[htp]
\subfloat[]{
\includegraphics[clip,width=\columnwidth]{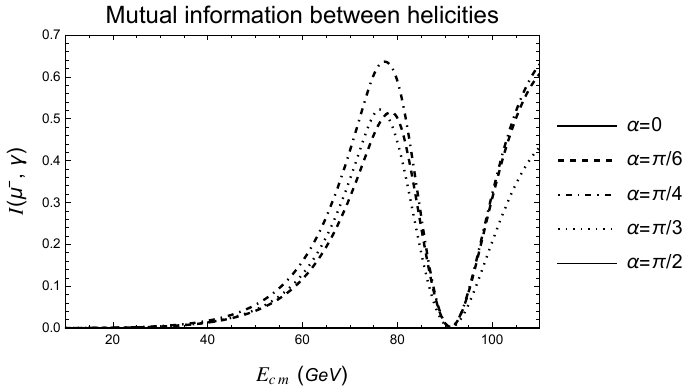}%
}

\subfloat[]{
\includegraphics[clip,width=\columnwidth]{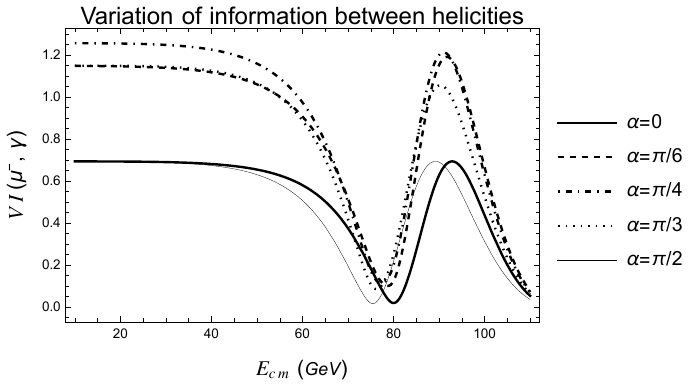}%
}
\caption{(a) and (b) plot the mutual information, $I(\mu^-,\gamma)$, and the variation of information, $VI(\mu^-,\gamma)$, between the muon's and witness photon's helicities versus $E_{cm}$.  $\alpha$ is the entanglement parameter.  $I(\cdot)$ quantifies the separability of the state whereas $V(\cdot)$ is similar to the correlation of helicities.}
\label{figmutvar}
\end{figure}

There are multiple confirmations of having the correct regularization of $V/(2T)$ as the total scattering cross-section ($\sigma_T$).  The reduced density matrices for $\mu^-,\gamma$ (see equations (\ref{rhoh}), (\ref{rhomom}), (\ref{rhogam})) are all Hermetian and have a trace of one.  Using equation (\ref{rhomom}), the conditional expectation value of the muon's momentum is $|\vb*l_1|$.
Using equation (\ref{rhogam}), the expectation value of the muon's helicity, $E(\lambda) = tr(\sigma_{z,\gamma}\ \rho^f_\lambda)$, is zero if the interaction occurs at low center of mass energy wherein the photon channel dominates over the $Z$-boson channel.  This is because electromagnetic interactions are invariant under parity.  The $Z$-boson and photon interference occurs at about $30\ GeV$.  Figure \ref{fighelicity}$(a)$ plots the conditional expected helicity for different entanglement parameters, $\alpha$. The curve with $\alpha = 45^\circ$ corresponds to having initial electron-positron pairs with equal weights in terms of helicities (see equation (\ref{initial})). At the $Z$-boson mass, all of the curves converge to the expected helicity
$E(\lambda) = -.21$ in units of $\hbar/2$.  In Figure \ref{fighelicity}$(b)$, the helicity entropy (see equation (\ref{slambda})) is flat at low energies and has a common value for the different $\alpha$'s at the $Z$-boson mass of $91.2\ GeV$.  As the magnitude of the expected helicity rises, the helicity entropy falls.

Figure \ref{figmutvar} plots both the mutual information, $I(\mu^-,\gamma)=S^{EE}_{\lambda} + S^{EE}_{\gamma} - S^{EE}_{\mu^-\gamma}$, and the variation of information, $VI(\mu^-,\gamma) = S^{EE}_{\mu^-\gamma} - I(\mu^-,\gamma)$,  between the muon's and witness photon's helicities.  $S^{EE}_\gamma$  is calculated from $tr_{\mu^-}(\rho^f_{\mu^-\gamma})$ and equation (\ref{rhoh}), not equation (\ref{redg}).  The final mutual information between helicities in Figure \ref{figmutvar}(a) is zero for $\alpha=0,\ \pi/2$.  At the latter entanglement parameter values, the initial state (see equation (\ref{initial})) is separable and its mutual information is zero.  Therefore, it appears appropriate to qualify the mutual information as a measure of separability of the state instead of as similar to correlation.  On the other hand, the variation of information is similar to the entanglement entropy and correlation of helicities (cf. Figure \ref{figmutvar}(b) and Figure \ref{helicities}).  The decrease in variation of information or distance between partitions is equivalent to both the increase in correlation or decrease in entropy.  Also, for the initial state, the correlation and variation of information for the electron's and photon's helicities are $-1$ and $0$, respectively.  Hence, it appears appropriate to interpret $VI(\cdot)$ as similar to correlation.

\section{Discussion}

Divergences are a common appearance in density matrix calculations for decay and scattering processes.  In section \ref{fdensity}, a Lorentz invariant regularization ansatz for a density matrix of an arbitrary interaction is given in equation (\ref{genreg}).  The correctness of the regularization relies on obtaining a density matrix of trace one and the expected helicities of final particles.  Also, this regularization provides the conditional density matrix given that the interaction occurs.  For the scattering of two particles and a decay process, the inverse time divergence represents the scattering rate and total decay width, respectively.  However, the ansatz could be relaxed.  For example, consider a muon decay.  An appropriate time regularization would be $-(1-e^{-t/\tau})/\Gamma$ where $t$ is the time and $\tau$ is the muon's lifetime.  After tracing over the muon, the final density matrix would be a direct sum of the probability that the muon does not decay ($e^{-t/\tau}$) and a matrix of final states for the muon decay.  The regularization ansatz given in this paper is for $t\rightarrow \infty$, which is the $S$-matrix formalism (cf. \cite{shiva2} and \cite{lello}).
 
This work may be extended by allowing the $\mu^-,\mu^+$ pair to decay into $e^-\bar \nu_e, e^+\nu_e$ with photon emission.  The mutual information of the latter final electron-positron pair would be diminished by the coherence length, i.e. Bremsstrahlung. Recently \cite{toumbas} has modeled soft photon emission, having the appearance of an unregularized area factor  $V/(\upsilon_{12}T)$ in the entanglement entropy.  This unregularized area should be interpreted as the total scattering cross section with photon emission in the final states.  

Further, given the regularization ansatz in this work, top quark pair production at the LHC may be investigated by calculating finite density matrices and finite mutual information between degrees of freedom such as spins.  The top quark is very suitable for study because it decays before hadronization, meaning the daughter leptons provide entanglement information  about the top quark pair.  Recently, Atlas has seen the first entanglement in a top quark pair \cite{ATLAS:2023}.   \cite{aoude} theoretically investigated the entanglement in top quark pairs with linear and quadratic levels of six-dimensional operators.  These operators are found to suppress the entanglement in the top quark pair.

\begin{appendix}

\section{Feynman amplitude and numerical constants}\label{feynamp-appendix}

\begin{figure}[ht]
\bc
\includegraphics[width=.7\textwidth]{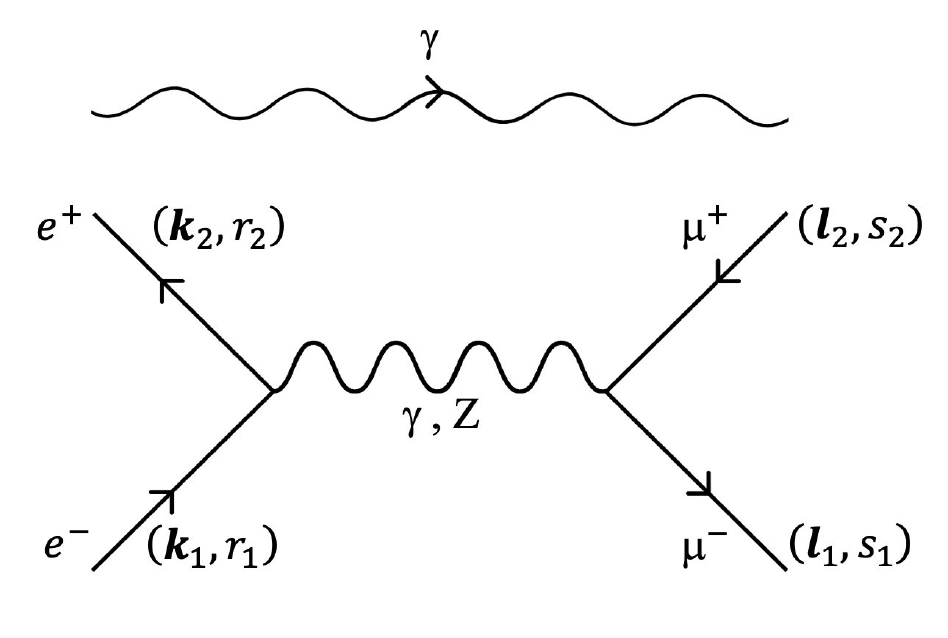}
\caption{ The Feynman  diagram $e^- e^+ \rightarrow \gamma,Z \rightarrow \mu^- \mu^+$ occurs in the presence of a witness photon that is entangled with the initial electron-positron pair. A photon and $Z$ boson are the propagators.    The momentum-helicity pairs for the electron and positron are  $(\vb*k_i, r_i), i=1,2$.  The momentum-helicity pairs for the muon and antimuon are  $(\vb*l_i, s_i), i=1,2$.}
 \label{FeynmanDiagram}
\ec
\end{figure}

$e^-e^+\rightarrow \mu^-\mu^+$ has both a photon channel and a $Z$ boson channel (see Figure \ref{FeynmanDiagram}).  $(k_1,k_2)$ and $(l_1,l_2)$ are the initial and final four momenta pairs, respectively.  The scattering process has a net center of mass energy $E_{cm}=\sqrt{s}=\sqrt{( k_1 + k_2)^2}$. The Higgs boson channel is ignored assuming $E_{cm}$ is not close to the Higgs mass.        $r_i,s_i$ are the leptonic helicities taking values $(R,L)$.  $u(\vb*k_1;r_1)$ and $u(\vb*l_1;s_1)$ are the spinors while $\upsilon(\vb*k_2;r_2)$ and $\upsilon(\vb*l_2;s_2)$ are the antispinors.   

The witness photon in Figure \ref{FeynmanDiagram} does not participate in the scattering.  The Feynman amplitude for the process $e^-e^+\rightarrow \gamma,Z \rightarrow \mu^-\mu^+$ \cite{peskin} is given by 
\begin{align}\label{feynman}
 &\mathcal{M}^{s_1,s_2}_{r_1,r_2}\equiv \mathcal{M}^{\vb*l_1,s_1;\ \vb*l_2,s_2}_{\vb*k_1,r_1;\ \vb*k_2,r_2 }=
 \frac{e^2}{s}\ \bar u(\vb*l_1;s_1)
   \gamma_\alpha \upsilon(\vb*l_2;s_2)
   \
   \bar \upsilon(\vb*k_2;r_2) \gamma^\alpha u(\vb*k_1;r_1) \ + \nn\\
   &\indent\indent \frac{G\ m_Z^2}{2\sqrt{2}(s-m_Z^2+im_Z  \Gamma_Z)}* 
   \bar u(\vb*l_1;s_1)
   \gamma_\alpha (b-\gamma_5)\upsilon(\vb*l_2;s_2)*\nn\\
   &
\indent\indent\indent
   (g^{\alpha\beta} - \frac{k^\alpha k^\beta}{m_Z^2})*\bar \upsilon(\vb*k_2;r_2) \gamma_\beta (b-\gamma_5) u(\vb*k_1;r_1).
\end{align}
 The  term $\dfrac{k^\alpha k^\beta}{m_Z^2}$ in the $Z$ boson propagator does not contribute since the electron and muon are assumed massless where the total energy is $E_{cm} = \sqrt{s}\geq10\ GeV$ in sections \ref{fdensity} and  \ref{QIS}.  $\alpha=\dfrac{e^2}{4\pi}=\dfrac{1}{137.035999084}$ and $G=1.1663789\text{x}10^{-5}\ GeV^{-2}$ are the fine structure and Fermi coupling constant, respectively.  $b=1-4\sin^2 \theta_W$ where the weak mixing angle relates the vector boson masses via $\cos \theta_W = \dfrac{m_W}{m_Z}=\dfrac{80.377}{91.1876}=.88145$ \cite{pdg1}.   The total decay width of the $Z$-boson is $\Gamma_Z = 2.4955\ GeV$ \cite{pdg2}.  Keeping the decimal places allows for the density matrix to have a trace close to one.

\subsection{Feynman amplitudes in helicity basis}\label{feynamps}

Using equation \ref{feynman} above, the four non-zero helicity amplitudes are the following, where $\theta$ is the muon's scattering angle with respect to the electron's momentum in the center of mass frame.
\begin{align*}
\mathcal{M}_{R,L}^{R,L}&=  e^2(1+\cos\theta) + \frac{G\ m_Z^2 s}{2\sqrt{2}(s-m_Z^2+im_Z \Gamma_Z)}\  (b-1)^2(1+\cos\theta)\\
\mathcal{M}_{L,R}^{L,R}&= e^2(1+\cos\theta) + \frac{G\ m_Z^2 s}{2\sqrt{2}(s-m_Z^2+im_Z \Gamma_Z)}\  (b+1)^2 (1+\cos\theta)\\
\mathcal{M}_{R,L}^{L,R}&= e^2(1-\cos\theta) + \frac{G\ m_Z^2 s}{2\sqrt{2}(s-m_Z^2+im_Z \Gamma_Z)}\  (b^2-1)(1-\cos\theta)\\
\mathcal{M}_{L,R}^{R,L}&= e^2(1-\cos\theta) + \frac{G\ m_Z^2 s}{2\sqrt{2}(s-m_Z^2+im_Z \Gamma_Z)}\  (b^2-1
)(1-\cos\theta)\\
\end{align*}
The first term in each amplitude above is from the photon channel.  The second term is from the $Z$-boson channel.  Unlike electromagnetism, the weak channel violates parity, introducing forward-backward asymmetry.

\subsection{Density matrix of muon and witness photon}\label{rhomug-appendix}
The density matrix of the muon's and witness photon's helicities is the following.

$\rho_{\mu^-\gamma}^f$ = $\dfrac{1}{\sigma_\mu} \dfrac{(\hbar c)^2}{12\pi s} *$\\\\

\begin{blockarray}{(cccc)}
  $\sin^2\alpha\ |b_3|^2$ & $\dfrac{ \sin2\alpha}{4}e^{i\beta}(b_1^*b_3)$ & 0 & 0\\
   $\dfrac{ \sin2\alpha}{4} e^{-i\beta} (b_1 b_3^*)$ & $\cos^2\alpha\ |b_1|^2$ & 0 & 0 \\
   0 & 0 & $\sin^2\alpha\ |b_2|^2$ & $\dfrac{ \sin2\alpha}{4}e^{i\beta}(b_3^*b_2)$ \\
   0 & 0 & $\dfrac{ \sin2\alpha}{4}e^{-i\beta}(b_3b_2^*)$ & $\cos^2\alpha\ |b_3|^2$
\end{blockarray}

The diagonal matrix elements of $\rho^f_{\mu^-\gamma}$ from the top left corner to the bottom right corner have the following order: $|R\ua \ra\la R\ua |, |R\da \ra\la R\da |,\\ |L\ua \ra\la L\ua |, |L\da \ra\la L\da |$.  ($R,L$) refers to a muon being right-handed or left-handed. ($\ua,\da$) refers to the witness photon being right-handed or left-handed.  The complex numbers $b_1,\ b_2,$ and $b_3$ above are the Feynman amplitudes $\mathcal{M}_{R,L}^{R,L},\ \mathcal{M}_{L,R}^{L,R},$ and $ \mathcal{M}_{L,R}^{R,L}$, respectively at $\theta=\pi/2$.  See appendix \ref{feynamps}.  e.g., the matrix element $\la R \ua| \rho^f_{\mu^-\gamma}|R \da\ra$ is $\dfrac{1}{\sigma_\mu} \dfrac{(\hbar c)^2}{12\pi s}\dfrac{ \sin2\alpha}{4}e^{i\beta}(b_1^*b_3)$.      

Notice $\rho^f_{\mu^-\gamma}$ has a trace of one and is hermetian.  The reduced density matrices for the witness photon and muon are $\rho_\gamma^f= tr_{\mu^-}(\rho_{\mu^-\gamma}^f)$ and $\rho_{\mu^-}^f =  tr_\gamma(\rho_{\mu^-\gamma}^f)$, respectively.  These reduced density matrices ($\rho^f_{\mu^-\gamma},\ \rho_{\mu^-}^f,\ \rho_\gamma^f)$ assume a $\mu^-\mu^+$ final state for the electron-positron scattering.

\end{appendix}

\setcounter{secnumdepth}{0}

\section{Acknowledgements}
Thanks to Ms$.$ Haiyu Zhu for initial work on this project.  GG and HZ were funded by a Colgate University Research Council grant.    

\section{Competing Interests}
Competing interests: The authors declare there are no competing interests.

\end{document}